\documentclass[authoryear,preprint,12pt]{elsarticle}

\usepackage{amssymb}
\usepackage{epsfig}

\begin{document}

\begin{frontmatter}



\title{The First  High-Precision Radial Velocity Search for Extra-Solar Planets}


\author{Gordon A. H. Walker}

\address{1234, Hewlett Place, Victoria  BC, V8S 4P7, Canada}

\ead{gordonwa@uvic.ca}

\begin{abstract}
The reflex motion of a star induced by a planetary companion is  too small to detect by photographic astrometry. The apparent discovery in the 1960s of  planetary systems around certain nearby stars, in particular Barnard's star, turned out to be spurious. Conventional stellar radial velocities determined from photographic spectra at that time were also too inaccurate  to detect the expected reflex velocity changes. In the late 1970s and early 1980s, the introduction of solid-state,  signal-generating detectors and absorption cells to impose wavelength fiducials directly on the starlight, reduced radial velocity errors to the point where such a search became feasible. Beginning in 1980, our team from UBC introduced an absorption cell of hydrogen fluoride gas in front of the CFHT  coud\'{e} spectrograph and, for 12 years,  monitored the radial velocities of some 29  solar-type stars. Since it was assumed that extra-solar planets would most likely resemble Jupiter in mass and orbit, we were awarded only three or four two-night observing runs each year. Our survey highlighted  three potential planet hosting stars,  $\gamma$ Cep (K1 IV), $\beta$ Gem (K0 III), and $\epsilon$ Eri (K2 V). The putative planets all resembled Jovian systems with periods and masses of:  2.5 yr and  1.4 $M_{J}$, 1.6 yr and 2.6 $M_{J}$, and 6.9 yr and 0.9 $M_{J}$, respectively. All three were subsequently confirmed from more extensive data by the Texas group led by Cochran and Hatzes who also derived the currently accepted orbital elements. 

None of these three systems is simple. All 5  giant stars and the supergiant in our survey proved to be intrinsic velocity variables. When we first drew attention to a possible planetary companion to  $\gamma$ Cep in 1988 it was classified as a giant,  and there was the possibility that its radial velocity variations and those of $\beta$ Gem (K0 III) were intrinsic to the stars. A further complication for $\gamma$ Cep was the presence of an unseen secondary star in an orbit with a period initially estimated at some 30 yr. The implication was that the planetary orbit might not be stable, and a Jovian planet surviving so close to a giant then seemed improbable. Later observations by others showed the stellar binary period was closer to 67 yr, the primary was only a sub-giant and a weak, apparently synchronous chromospheric variation  disappeared. Chromospheric activity was considered important because $\kappa^{1}$ Cet, one of our program stars, showed a significant correlation of its radial velocity curve with chromospheric activity. 

 $\epsilon$ Eri is a young, magnetically active star with spots making it a  noisy target for radial velocities. While the signature of  a highly elliptical orbit ($e=0.6$)  has persisted for more than 3 planetary orbits, some feel that even more extensive coverage is needed to confirm the identification  despite an apparent complementary astrometric acceleration detected with the Hubble Space Telescope.

We confined our initial analyses of the program stars to looking for circular orbits. In retrospect, it appears that some 10\% of our sample did in fact have Jovian planetary companions in orbits with periods of years. 

\end{abstract}


\end{frontmatter}

\section{The very  idea of searching for extra-solar planets in the 1970s}
\label{}

It is quite hard nowadays to realise the atmosphere of skepticism and indifference in the 1980s to proposed  searches for extra-solar planets. Some people felt that such an undertaking was not even a legitimate part of astronomy. It was against such a background that we began our precise radial velocity (PRV) survey of certain bright solar-type stars in 1980 at the Canada France Hawaii 3.6-m Telescope (CFHT). Up to that time, the only serious searches for extra-solar planets had been astrometric studies that looked for long term periodic displacements of stars on the sky caused by unseen companions. Unfortunately, the tiny angular displacements were much smaller than the proper motions and parallaxes for nearby stars and beyond the capability of ground-based astrometric telescopes using photographic emulsions.

For a circular planetary orbit, the semi-angular reflex displacement, $\theta$, of the parent star on the sky caused by a companion planet is:
\begin{equation}
\theta \approx \frac{R_{pl}M_{pl}}{~D~M_{st}} ~~~~ {\rm milli\mbox{-}arc sec},
\end{equation}
where $R_{pl}$ is the distance of the planet from its parent star in AU, $M_{pl}$ and $M_{st}$ are the planet and stellar masses in Jovian and solar masses, respectively. $D$ is the distance of the star from the Earth in parsecs. $\theta$ increases linearly with $R_{pl}$ (and as the $\frac{2}{3}$ power of the period). For the Sun-Jupiter system seen from a distance of 5 pc, $\theta$ would be $\sim$1 milli-arcsec and take nearly 12 years to complete a cycle. Considering that the average ground-based photographic image was spread over $>$1000 milli-arcsec, the challenge of detecting such a miniscule displacement over several decades is obvious.

Nonetheless, tentative extra-solar planetary discoveries were claimed for several nearby stars from astrometry with the Sproul Observatory 24 inch refractor, but the signals were close to the errors of measurement. The best known identifications were for Barnard's Star \citep{Kam69}. The orbital periods in all cases were close to multiples of 8 years, and \citet{Her73} demonstrated that the epochs of the `discontinuities' in the Sproul data coincided with times when the telescope primary lens had been removed for cleaning.

The amplitude of the periodic reflex variation of a planet-hosting star's radial velocity complements the astrometric variation. Again, for circular orbits, the amplitude of the stellar radial velocity change, $\delta RV$, would be:
\begin{equation}
\delta RV \approx \frac{30M_{pl}~{\rm sin}i}{(M_{st}~R_{pl})^{1/2}}~~~~ {\rm km~~ s^{-1}},
\end{equation}
where $i$ is the inclination of the orbital axis to the line of sight and the units are the same as those in equation 1. The radial velocity excursion is independent of the distance from the Earth and increases inversely as $\surd R_{pl}$. As a star, $\delta RV$ for the Sun as caused by Jupiter would be $\sim\pm$13 m s$^{-1}$ over 12 years when viewed at $i$=0. Unlike astrometry, radial velocities favor the detection of tightly bound planets.

In the early 1970s, quality radial velocities determined from photographic plates had typical errors of $\sim$1 km s$^{-1}$. The large error was in part due to the photographic process but, more seriously, to the effect of guiding errors at the spectrograph slit which cause the stellar spectrum to shift relative to the bracketting comparison arc lines. Obviously, an improvement of more than an order of magnitude coupled with a long term stability of years, was necessary  if radial velocities were to be useful in looking for extra-solar Jovian analogs.

The above comments also reflect the belief at the time that planetary systems beyond the solar system would most likely be similar to our own, a belief strongly bolstered by theoretical simulations even down to the Titus Bode relation \citep{Isa77}! No one predicted the populations of Jovian planets in few-day, or highly elliptical, orbits which eventually turned up. As a result, early search strategies concentrated on long-term monitoring with observations spaced out over months and years. 


\section{The UBC PRV Planet Search}

Development of digital TV systems for astronomical spectroscopy began at UBC in 1969. These evolved from conventional cameras with electron-beam read-out to self-scanned silicon diode arrays \citep{Wal77}. Access to the horizontal McKellar coud\'e spectrograph fed by the 1.2-m Dominion Astrophysical Observatory (DAO) telescope provided a perfect laboratory in which to test our cameras \citep{Ric72}.

With conventional TV cameras, electron reading-beam jitter and beam-bending are serious problems in maintaining fidelity of stellar line positions in the digitised spectra. These errors were in addition to those from guiding - although the introduction of the Richardson (1966) image slicer significantly reduced the latter. Glaspey and Fahlman \citep[see][]{Wal73} demonstrated that one could use telluric water vapor lines to correct for all of these shifts, raising the possibility of achieving the true radial velocity precision inherent in the original data. The use of telluric lines was not a new idea \citep[e.g.][]{GG73}, but it was particularly appropriate to the case of digital spectra of wide dynamic range and high signal to noise, and I realised that it raised the possibility of sufficient precision to look for extra-solar planets.

The introduction of  self-scanned diode arrays eliminated all of the problems with electron beams and made the planet search a serious possibility. Bruce Campbell who joined me as a Post-Doctoral Fellow picked up on the idea and took it a step further by suggesting a captive gas. On the recommendation of Alexander Douglas and Gerhard Herzberg of the Canadian National Research Council Herzberg Institute of Astrophysics, hydrogen fluoride (HF) was chosen because the 3--0 vibration band has a well spaced comb of lines with no confusion from isotopic shifts, the lines are of similar natural width to those in the stellar spectra, and there are no telluric lines contaminating the $\lambda$8700 region. 

Bruce and I  \citep{Cam79} set up an absorption cell just ahead of the McKellar spectrograph slit on 27 December 1978 to try the first `stellar' observations. Frankly, it was quite unsafe. HF is highly corrosive and toxic. The cell had to be heated to 100 C to prevent the HF from polymerising and the cell windows being plexiglass (HF attacks glass) warped with the heat. Nonetheless, we took a series of exposures of the Sun with the telescope mirror covers closed - enough light got through the gaps between the covers to give us a good signal from the 1024 Reticon diode array. 

It worked! Thanks to the small spectral shift program developed by Fahlman and Glaspey we were able quickly to analyse the spectral series and demonstrate that HF fiducial lines could in principle reduce radial velocity errors by the necessary order of magnitude from those being determined photographically. One important aspect of the captive gas technique (as opposed to the use of telluric lines) is that stellar spectra can also be taken with the absorption cell out of the light beam to provide a template spectrum against which to determine differential velocities. Further, unlike the situation with telluric lines, spectra of the captive gas can be taken independently with a continuum source to allow the HF lines to be carefully aligned and removed from the stellar spectra.  Examples of  stellar spectra with and without the HF lines are shown in Figure 1 \citep{Cam88}. 

\citet{Bec76} was probably the first to use a captive gas, I$_{2}$, to determine wavelength shifts accurately in solar spectra \citep[see also][]{Koc84}. \citet{Mar92} later adopted I$_{2}$ for their very extensive PRV program.

It should be emphasised that our radial velocities were precise but {\it not} accurate, and only appropriate for measuring differential velocities or the accelerations of the target stars. The wavelengths of  stellar absorption lines depend on conditions in the stellar atmosphere in addition to the star's actual motion. There are upwelling velocities $\sim$1 km s$^{-1}$ of gasses hotter than their surroundings as well as cooler descending material, leading to the well known granulation pattern on the Sun. We had to assume that the average, weighted surface velocity remained constant (a doubtful assumption in many cases). The Earth itself is in revolution at 30 km s$^{-1}$ about the solar system barycentre and  measured velocities must be corrected for this effect, which can amount to tens of km s$^{-1}$ between observing runs and changes of several m s$^{-1}$ during a single exposure. We had to accurately  correct for the latter by estimating the mean time of the exposure weighted by the flux rate of starlight reaching the detector.  

The wavelength region covered by the HF lines (Figure 1) included one of the red Ca II triplet lines at $\lambda$8662. The strength of this line is sensitive to chromospheric activity. Such activity introduces reversals of variable intensity within the Ca II lines. We considered this an important bonus because  stellar granulation patterns might be affected by magnetic cycles and starspots thereby introducing apparent changes in radial velocity and we would be in a position to calibrate the effect if it existed. The period of Jupiter's orbit and the sunspot cycle are similar for instance.

Bruce become a staff astronomer at CFHT in 1979 while there he built a safer version of the HF cell with sapphire windows for use with the CFHT coud\'{e} spectrograph. The latter was a duplicate of the high dispersion McKellar spectrograph, and we, together with Stephenson Yang,  used the HF system at CFHT over the following twelve years for PRV. The detector, by then an 1872 diode array, had a correlated double sampling system \citep{Wal85} based on a design by \citet{Gea79} which reduced read-out noise to the point where  signal to noise was dominated by photon statistics. Full descriptions of the technique and analyses can be found  in  \citet{Yan82} and \citet{Cam88}.  To achieve a precision of $\sim$10 m s$^{-1}$ (equivalent to 0.006 of a 15$\mu$m diode spacing!) several subtle effects had to be calibrated, in particular the impact of convolving the illumination pattern at the slit with the HF line profiles. Ultimately, there also turned out to be small but significant systematic off-sets in the overall radial velocities between observing runs which could be corrected iteratively only at the end of the whole program.

\section{The UBC PRV Program}

\citet{Yan82} carried out the first successful initial program in 1980 using the HF absorption cell on the CFHT  to determine the low amplitude velocity variations of the $\delta$ Scuti star $\beta$ Cass. This was before the high reflectance coud\'{e} mirror train was installed. The wavelengths of the HF lines, close to $\lambda$8700,  coincide with a region of reduced  reflectivity for aluminum resulting in only some 10\% overall transmission, a situation soon remedied by the introduction of enhanced silver  coatings. This preliminary exercise gave us experience and helped to improve the associated reduction programs. The results for $\beta$ Cass demonstrated to the satisfaction of successive telescope time assignment committees that the HF PRV technique was indeed viable. 

There were two parts to the PRV observing list. Five bright giant and one super-giant K stars were included because they were regulary observed around the world as radial velocity standards.  The basic observing list consisted of 23 solar main sequence and sub-giant stars visible from CFHT \citep{Wal95}. Known spectroscopic binaries, rapid rotators and stars earlier than F5 were excluded. The stars are listed in Table 1. $\pi$ is the parallax in asec, the I magnitude corresponds most closely to the wavelength of the HF spectral region, and the total number of observations made of each star is in the final column.


\begin{table}
\caption{The UBC PRV Program Stars}
\begin{tabular}{rrcclrc}
\label{TableObs}
\\ \hline \hline
HR~ &HD ~ & name  &$\pi$ &Sp. type  &I ~&obs \\
\hline
Giants&&&&&&\\
617   &12929 & $\alpha$ Ari&0.049&K2 IIIab &0.52& 25\\
1457& 29139  & $\alpha$ Tau& 0.054& K5 III& --1.29 & 32\\
2990& 62509  & $\beta$ Gem& 0.094 & K0 IIIb&--0.12 &20\\
3748&  81797 & $\alpha$ Hya& 0.022 & K3 II--III&0.15 & 32\\
5340& 124897  & Arcturus&   0.097     &K1.4 III& --1.70 & 42\\
8308&  206778 & $\epsilon$ Peg& 0.006 &K2 Ib&  0.57   &21\\
Dwarfs&&&&&&\\
509 & 10700 & $\tau$ Cet & 0.287 & G8 V & 2.41 & 68 \\ 
937 & 19373 & $\iota$ Per & 0.092 & G0 V & 3.25 & 46 \\
996 & 20630 & $\kappa^1$ Cet & 0.108 & G5 Vvar & 3.95 & 34 \\
1084 & 22049 & $\epsilon$ Eri & 0.304 & K2 V & 2.54 & 65 \\
1325 & 26965 & $o^2$ Eri & 0.209 & K1 V & 3.29 & 42 \\
2047 & 39587 & $\chi^1$ Ori & 0.104 & G0 V & 3.61 & 38 \\
2943 & 61421 & $\alpha$ CMi A & 0.292 & F5 IV-V & --0.27 & 98 \\
3775 & 82328 & $\theta$ UMa & 0.068 & F6 IV & 2.47 & 43 \\
4112 & 90839 & 36 UMa & 0.077 & F8 V & 4.08 & 56 \\
4540 & 102870 & $\beta$ Vir & 0.104 & F9 V & 2.86 & 74 \\
4983 & 114710 & $\beta$ Com & 0.124 & F9.5 V & 3.46 & 57 \\
5019 & 115617 & 61 Vir & 0.117 & G5 V & 3.82 & 53 \\
5544 & 131156 & $\xi$ Boo A & 0.156 & G8 V & 3.75 & 58 \\
6401 & 155886 & 36 Oph A & 0.188 & K0 V & 3.99 & 26 \\
6402 & 155885 & 36 Oph B & 0.188 & K1 V & 4.02 & 35 \\
7462 & 185144 & $\sigma$ Dra & 0.177 & K0 V & 3.66 & 56 \\
7602 & 188512 & $\beta$ Aql & 0.077 & G8 IV & 2.59 & 59 \\
7957 & 198149 & $\eta$ Cep & 0.076 & K0 IV & 2.27 & 58 \\
8085 & 201091 & 61 Cyg A & 0.294 & K5 V & 3.54 & 50 \\
8085 & 201092 & 61 Cyg B & 0.294 & K7 V & 3.54 & 34 \\
8832 & 219134 & & 0.146 & K3 V & 4.23 & 32 \\
8974 & 222404 & $\gamma$ Cep &0.068 & K1 IV&1.93  & 69 \\
\hline
\end{tabular}
\end{table}

We applied twice each year for telescope time and were generally assigned four pairs of nights per year - although one year we received none in the first six months and the TAC forced us eventually to drop the giants from the program and reduced us to three pairs of nights per year. With the mindset that we were waiting for planetary signals with characteristic signals of a decade, it really was tedious  because there could be no obvious results from any one observing run and, perhaps more seriously, no publications to nourish research funds. Nonetheless, we did persist for 12 years and, it seems, made some exo-planetary discoveries in the process - but it took some time for these to dawn because none was a simple case.

\subsection{$\kappa^{1}$ Ceti and the $\lambda$8662 Ca II chromospheric index. }

As mentioned earlier, the chromospheric activity of the individual program stars could be quantified from changes in the equivalent width of the chromospherically-sensitive, singly ionised calcium triplet lines one of which, at  $\lambda$8662, was included in our spectra. Details of how the index was formed are given in \citet{Lar93}. Basically, differences from a mean Ca II profile derived from all of the spectra were expressed as $\Delta$EW$_{866.2}$ in picometer units (1 pm = 10$^{-2}$ \AA).

We expected that there might be a  correlation between $\Delta$EW$_{866.2}$ and excursions in the PRV curve.  Our observations of $\kappa^{1}$ Cet appeared to support this idea in dramatic fashion as one can see  in Figure 2, where  $\Delta$EW$_{866.2}$ and the PRV values are scaled and plotted on the same diagram. The errors on the values of $\Delta$EW$_{866.2}$ are similar to the size of the points while for the PRV values they are $\sim\pm$13 m s$^{-1}$.  The two quantities track each other surprisingly well, particularly through the, as yet unexplained, S-shape excursion in 1988. 

In Figure 3 the individual PRV and $\Delta$EW$_{866.2}$ are plotted against each other. Despite the increasing scatter at large PRV values, there is clearly a significant correlation which was to lead to legitimate doubts about the apparent planetary PRV signatures for some chromospherically active stars and whether a correction based on the $\kappa^{1}$ Cet correlation was valid.  Ultimately, we did not apply any such correction because there was no clear case for any other star.   

Later observations of $\kappa^{1}$ Cet by \citet{Ruc04} and \citet{Wal07} showed that not only is the star chromospherically active, but that it is spotted and rotates differentially and the chromospheric activity is modulated at a rotational period of 9.3 d.  They found no obvious rotational modulation of radial velocity and pointed out that the $\sigma_{RV}$=24.4 m s$^{-1}$ found over 11 yr by \citet{Cum99} would allow a  Jovian planet in a tidally locked orbit to remain undiscovered. One might also speculate that the significant PRV and $\Delta$EW$_{866.2}$ excursion seen in 1988 lasting some 30 d might have been induced by close periastron passage of a planet in a highly eccentric orbit  by the mechanism suggested by \citet{Cun00} and for which \citet{Shk05} later found evidence in HD 179949 and $\upsilon$ And.

For good reason, current PRV searches try to avoid chromospherically active stars, but that does not mean they do not harbour planets!

\subsection{The Giants}

From initial PRV results one had to conclude that the so-called standards were all variable in radial velocity while most of the program stars were constant! While the range of  variability in the giants was not serious enough to prevent their use as standards for conventional photographic radial velocities, their radial velocities varied between 30 and 300 m s$^{-1}$ with periods $\sim$1 year. We published these results \citep{Wal89} after 5 years to point out that the yellow giants were a new class of radial velocity variable. In particular, Arcturus showed a range of periods from days to nearly two years \citep{Irw89}.  Because of the sizes of the giant stars,  it was assumed that their long  rotation  periods  contributed  part of the long-term variability.

\subsubsection{$\beta$ Gem}

As PRV  results for the giants accumulated, the repitition of some variability bore the signature expected from perturbation by  Jovian companions in circular orbits, particularly in the case of $\beta$ Gem \citep{Ana93} where we eventually extended the CFHT PRV series to mid-1993 with HF observations from the DAO. For $\beta$ Gem, some eight cycles of a 584.7 day period emerged with an amplitude of 46 m s$^{-1}$ that could correspond to a companion with $M$sin$i$=2.5 M$_{J}$ (see Figure 4). 

\citet{Hat93} had also reported variable radial velocities for three K giants, one of which was Pollux(=$\beta$ Gem) and, in 2006, \citet{Hat06} published the 25 year radial velocity curve shown in Figure 4. The velocities include the CFHT and DAO HF PRV values and extended observations from the  McDonald and Thuringia State, Observatories.  There is a clear 589.6 day period with an amplitude of 41 m s$^{-1}$ which, for a 1.7 solar-mass primary  corresponds  to M$_{pl}$sin$i$=2.3 M$_{J}$ with the planet in a nearly circular orbit ({\it e} = 0.02) and a semi-major axis of 1.6 AU. The persistance of  the  589.6 day  signal  ruled out pulsational or rotational variability of the star as plausible, alternate explanations. There seems little doubt about the reality of  this planetary identification.

 \subsection{The Main Sequence Stars}

After six years, we \citep{Cam88} published interim PRV results for 12 solar main sequence stars and 4 sub-giants. Unfortunately, our suggestion in that paper that statistically significant long-term trends for seven of the stars might represent `the tip of the planetary mass spectrum' only seemed to intensify  skepticism about the possibility of extra-solar Jovian planets.  $\gamma$ Cep and $\epsilon$ Eri were two of the candidates. 

Sometime later, Bruce Campbell, who had been unable to secure a permanent position in astronomy, left the project. Alan Irwin stepped in to help with the data analysis and we \citep[][]{Wal95} eventually published a final analysis for 22 stars based on the full 12 years of observations. (Published in Icarus because they had no page charges and by then we had exhausted our publication funds!) By that time, \citet{Wol92} had published their detection of terrestrial-mass planets orbiting the 6.2-ms pulsar PSR B1257 + 12, but it was just prior to the dramatic announcement of the Jovian-mass planet orbiting 51 Peg with a  4.23 d period by \citet{May95}. Ironically, the latter discovery was made without the benefit of  captive gas fiducials but relied on a highly stable spectrograph with a stable line spread function and comparison arc lines. 

There were several obvious conclusions from our twelve years of PRV observations. While we had detected two previously unknown spectroscopic binaries ($\gamma$ Cep and $\chi^{1}$ Ori) there was no evidence for brown dwarf companions to any of our stars similar to, or more massive than, that found by  \citet{Lat89}  associated with HD 114762. This apparent brown dwarf `desert' had turned up  in other studies \citep[e.g.][]{Duq91,Fis92}. Several stars such as $\tau$ Cet had shown surprisingly little change in radial velocity over the 12 years while others had proved much more noisy suggesting a higher level of surface activity.

In our 1995 paper, we struck a more conservative note than in 1988, concluding that there was no strong evidence for any Jovian planets in circular orbits with periods of years associated with any of our stars - a conclusion reinforced by the lack of detections by other groups.  However, we  did draw  attention to two candidates, $\gamma$ Cep and $\epsilon$ Eri, where the stellar velocity curves could indeed be interpreted as due to Jovian planets.  Neither  case was simple, which is why they remained in doubt until the excellent follow-up work of the Texas planet searchers led by Bill Cochran \citep{Wit06}. 

\subsection{$\gamma$ Cephei}

We had found \citep{Cam88} that $\gamma$ Cep was a single-line spectroscopic binary from a `large' systematic change in its radial velocity, but our observations covered only part of the binary orbit (see Figure 5). Subsequently, \citet{Tor07} was able to assign an orbital period of  67 yr  to the system from much more extended radial velocity measurements.  Recently, \citet{Neu07} actually imaged the faint second star with the aid of adaptive optics on the Subaru Telescope - it is likely an M4 dwarf. 

The interest to us was less in the stellar binary than in the 2.5 yr `ripple' superimposed on the velocity curve seen best perhaps in Figure 5  taken from \citet{Wal92} that shows the radial velocity curve from the first 11 years of the program. On removing a low-order-polynomial to account for the reflex motion of the primary caused the long period stellar binary orbit, there was an underlying periodicity from which \citet{Cam88} had suggested `Probable third body variation of 25 m s$^{-1}$ amplitude, 2.7 yr period, superimposed on a large velocity gradient'. With the 11 year curve, the periodicity in the residuals was much clearer, and the period was refined to 2.52 years. 

But was it a planet or an effect of the star's rotation? At the time, rotation seemed a slightly more plausible explanation because of a very weak almost synchronous chromospheric activity signal coupled with a misclassification of the primary as a giant. There was also the implausibility that there would be a stable planetary orbit in such a close stellar binary, and the then prevelant view that Jovian planets would not survive so close to such a large primary. Unfortunately, our best guess at a binary orbital period of $\sim$30 yr later proved too short \citep{Tor07}, and  \citet{Fuh03} established that the primary was a K1 IV sub-giant, not a giant. Both of these developments undercut any stellar origin for the periodic variations. Ultimately \citet{Hat03}, by  extending measurements of PRV  and  chromospheric activity  to 2002 at the McDonald Observatory, demonstrated that the 2.5 yr period in the residuals had repeated like clockwork and that there was in fact no long-term synchronous  chromospheric activity. This is shown dramatically in Figure 6 where the 22 years of differential radial velocities corrected for the velocity changes induced by the stellar binary are shown for each of the CFHT and three McDonald programs. Zero-point corrections were introduced between each of the four series for continuity. 

According to \citet{Tor07}, the M4 dwarf secondary star is in an eccentric 66.8  yr orbit ($e$ = 0.41) with a semimajor axis of 19 AU.  The minimum mass of the  planetary companion to the primary is 1.4 M$_{J}$. The planetary orbit  is only 9.8 times smaller than the orbit of the secondary star (the smallest ratio known among exoplanet host stars in multiple systems at this time), but it is stable if the two orbits are coplanar. 

So, arguably we detected the first exo-planet PRV signature for a Jovian planet in a quite unexpected system - a binary star, but  it was at a characteristically Jovian distance from its primary.

\subsection{$\epsilon$ Eri}

$\epsilon$ Eri is a young \citep{Sod89}, magnetically active star \citep{Gra95} which rotates twice as fast as the Sun.  It also has spots and the rotation is differential \citep{Cro06}. As a result, its integrated surface velocity changes make it a `noisy' target producing a considerable intrinsic scatter in radial velocity. Despite the `noise', there appeared to be a consistent $\sim$10 yr period in the UBC PRV, but the time baseline was too short to confirm this. From observations at McDonald, Lick and ESO observatories,  \citet{Hat00} extended coverage to 2000. The periodicity was still present, and they derived a period of 6.9 yr with an amplitude of  19 m s$^{-1}$ corresponding to a minimum planetary mass of  0.9 Jupiter in an eccentric orbit ($e$=0.6). 

Figure 7 shows the radial velocity curve extended to 2006  by \citet{Ben06}, who combined these values with three years of astrometric measurements of  milliarsecond precision taken with the HST. The noisiness of the radial velocity observations is obvious and  the astrometry covers less than half an orbit.

\section{Conclusions}

By 1993, when our PRV program ended, the HF technique had passed its prime. Our observing list was necessarily small because the 100 \AA~ coverage was an order of magnitude less than with I$_{2}$ vapor and  CCD arrays coupled to an  \'{e}chelle spectral format \citep[e.g.][]{Mar92,Wit06}.  Indeed,  \citet{May95} had even demonstrated that with sufficient care one could perhaps dispense with an absorption cell altogether.  The handling and preparation of the HF system was also dangerous. On the other hand, the HF system had been straight-forward to implement. The natural width and spacing of the HF lines had avoided the additional layer of processing necessary  with the sharp overlapping  I$_{2}$ lines. The photometric fidelity of the diode arrays was ideal, and we already had a `small-shift' progam in place as a basis for the analysis. Our early success in reducing PRV errors  had, we liked to think, encouraged other groups to press on.

As with any new level of precision, our PRV turned up unexpected results such as the variability of the yellow giants, the correlated variations of  the PRV and chromospheric activity in $\kappa^{1}$ Cet, and a paucity of close brown dwarf companions. These finds complicated interpretation of our results for the program stars. Above all,  a general skepticism still greeted any announcement of an exo-planet discovery. Nonetheless, three stars that we flagged as possible planet hosts, $\gamma$ Cep, $\beta$ Gem, and $\epsilon$ Eri  were later confirmed by the Texas group, albeit not everyone is yet convinced about  $\epsilon$ Eri b despite the complementary astrometric acceleration detected by \citet{Ben06}.  In simple terms, roughly some 10\% of our program stars turned out to have Jovian planets with orbital periods of years. This roughly matches the statistics from the current huge sample of stars being monitored by others  \citep[see for example][]{Mar00}, although each of our three candidates was unusual.  

Since our stars were only observed a few times per year, we had little chance of detecting closely bound systems like 51 Peg, but the results for  $\kappa^{1}$ Cet are tantalising in that respect.

The success of our extra-solar planet search depended on many people, but particularly on Bruce Campbell, Stephenson Yang, Alan Irwin and Ron Johnson who built successive low-noise versions of the Reticon camera systems. It is a great pleasure to be able to acknowledge my debt to them here.


\newpage

\begin{figure}
\begin{center}
\epsfig{figure=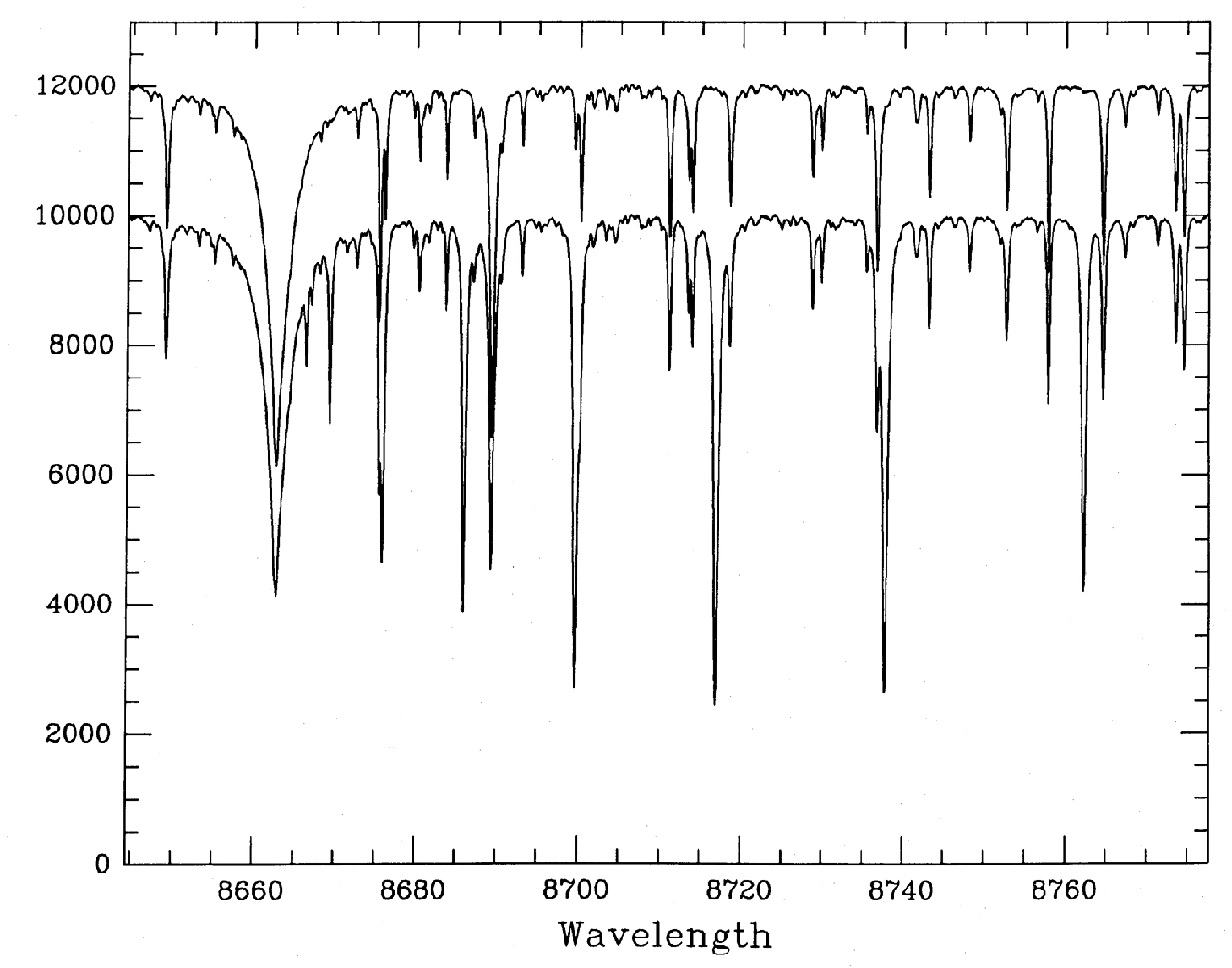,height=11cm,angle=0,clip=}
\end{center}
\caption{Two spectra of 61 Cyg A taken with the CFHT coud\'{e} spectrograph. The lower one was taken with the 80 cm hydrogen fluoride absorption cell in the beam and  shows the 3-0  HF band lines superimposed, while the upper one was taken without the HF cell. The broad stellar feature at $\lambda$8662 is one of the Ca II triplet lines the strength of which was used as a measure of chromospheric activity. \citep[Taken from][]{Cam88}}
\label{HF}
\end{figure}


\begin{figure}
\begin{center}
\epsfig{figure=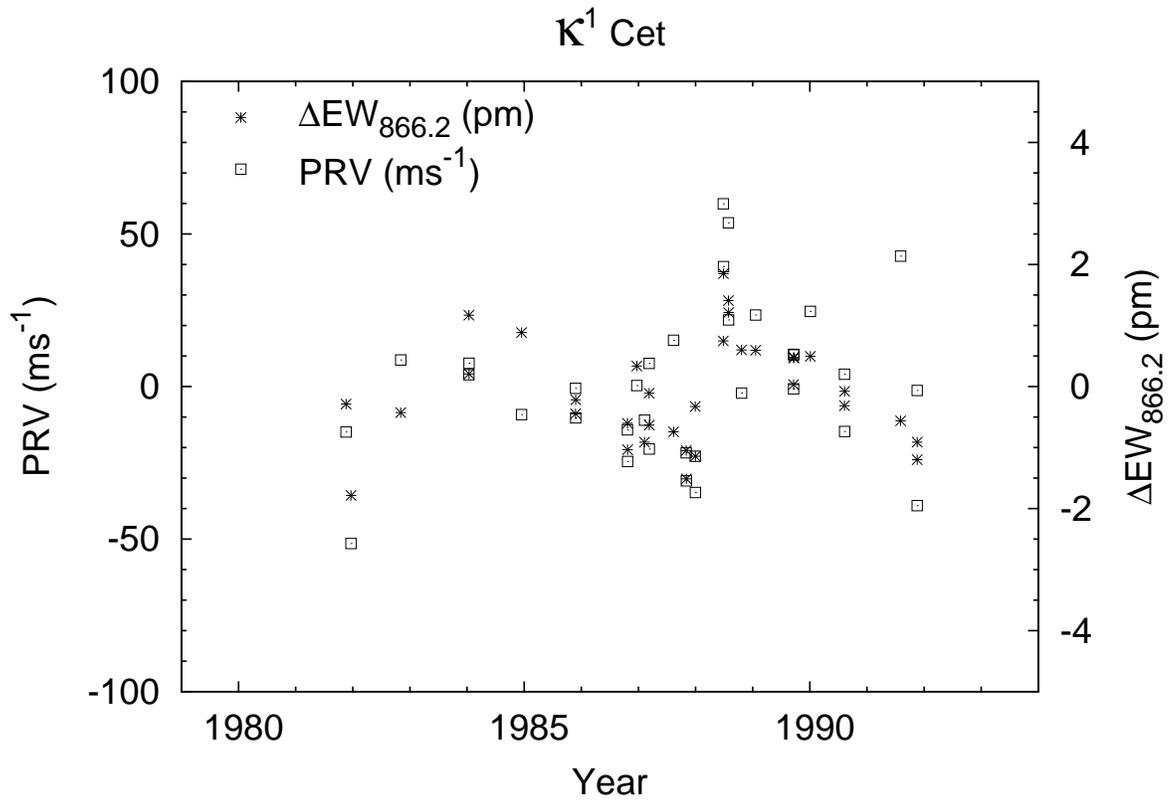,height=11cm,angle=0,clip=}
\end{center}
\caption{{This double plot demonstrates how  the precise radial velocities (PRV) of $\kappa^{1}$ Cet and its  chromospheric activity index ($\Delta$EW$_{866.2}$) varied in unison over eleven years.  The effect is particularly marked through the S-shaped change in 1988. The uncertainties in the $\Delta$EW$_{866.2}$ values are similar to the size of the points and the  uncertainties  in the PRV are $\sim\pm$13 m s$^{-1}$.  \citep[Adapted from][]{Wal95}}  
	\label{kCet}
}
\end{figure}

\begin{figure}
\begin{center}
\epsfig{figure=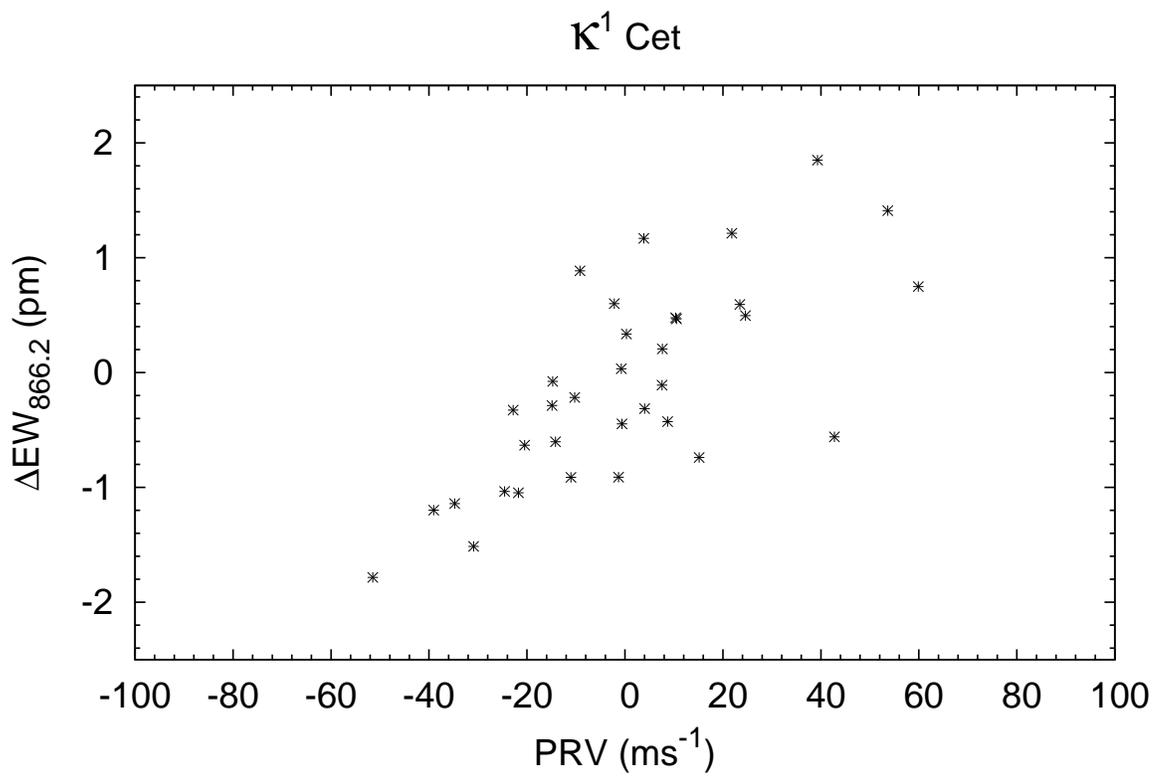,height=11cm,angle=0,clip=}
\end{center}
\caption{{The $\Delta$EW$_{866.2}$ values for  $\kappa^{1}$ Cet in Figure 2 plotted against PRV demonstrating a significant correlation. }  
	\label{corr}
}
\end{figure}

\begin{figure}
\begin{center}
\epsfig{figure=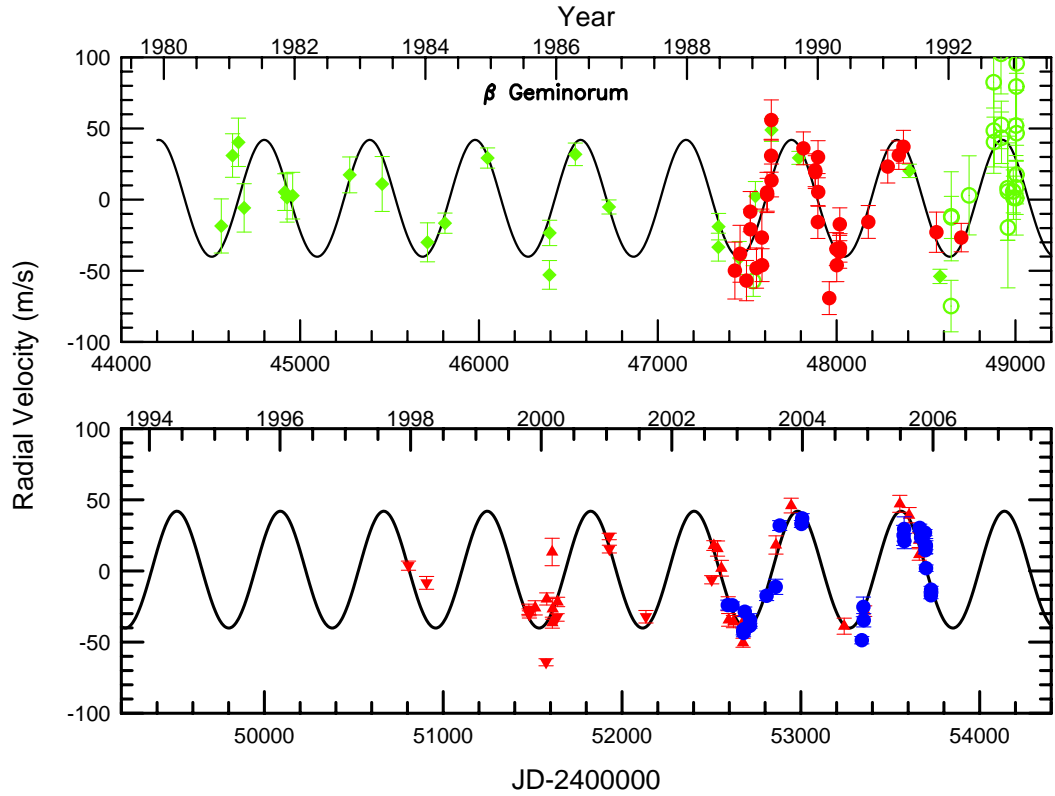,height=16cm,angle=270,clip=}
\end{center}
\caption{{Precise radial velocities for the K0 giant $\beta$ Gem taken over 25 years from 6 data sets, CFHT (green diamonds), DAO (red filled circles), three sets from the McDonald Observatory (green open circles, red inverted triangles and bluesquares), and the Thuringia State Observatory (red triangles). The solid line corresponds to a 589.6 day period orbit and a minimum planetary mass of 2.3 Jupiter and a nearly circular orbit ({\it e} = 0.02) with a semi-major axis of 1.6 AU.  \citep[Taken from][]{Hat06}}  
	\label{bGem}
}
\end{figure}

\begin{figure}
\begin{center}
\epsfig{figure=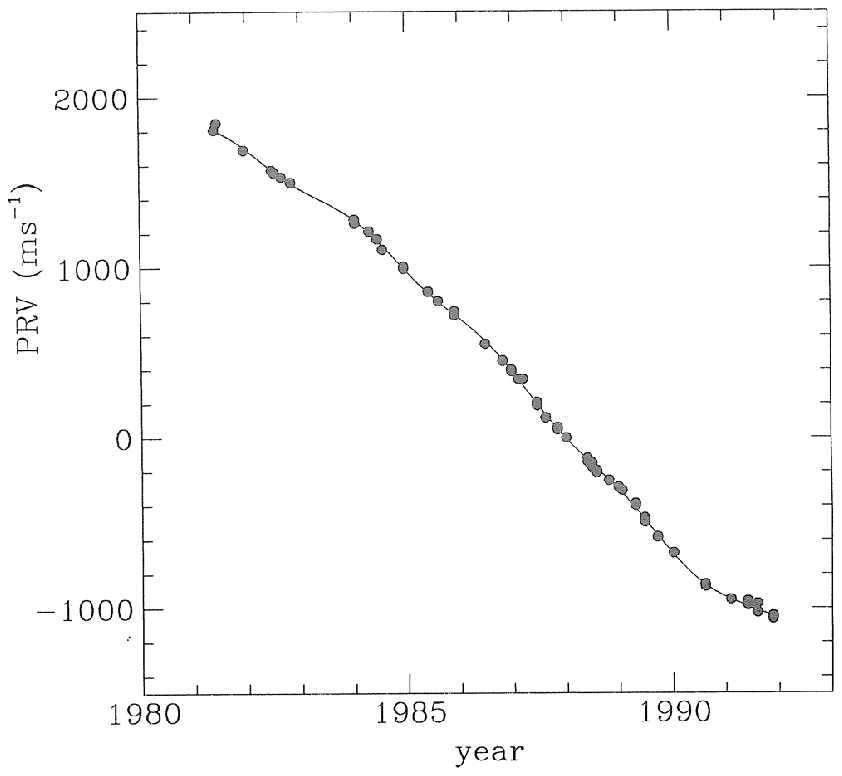,height=12cm,angle=-0.5,clip=}
\end{center}
\caption{The 11-year CFHT PRV curve for $\gamma$ Cep \citep{Wal92}. The overall change of $\sim$1 km s$^{-1}$ is caused by the reflex motion of  the primary induced by a much fainter secondary star \citep[later detected by][]{Neu07} in an orbit of decades. The ripple superimposed on the binary star velocity trend was first noted  by \citet{Cam88} and corresponds to reflex motion of the primary in response to a companion Jupiter-mass planet.  } 
\label{gCep}
\end{figure}

\begin{figure}
\begin{center}
\epsfig{figure= 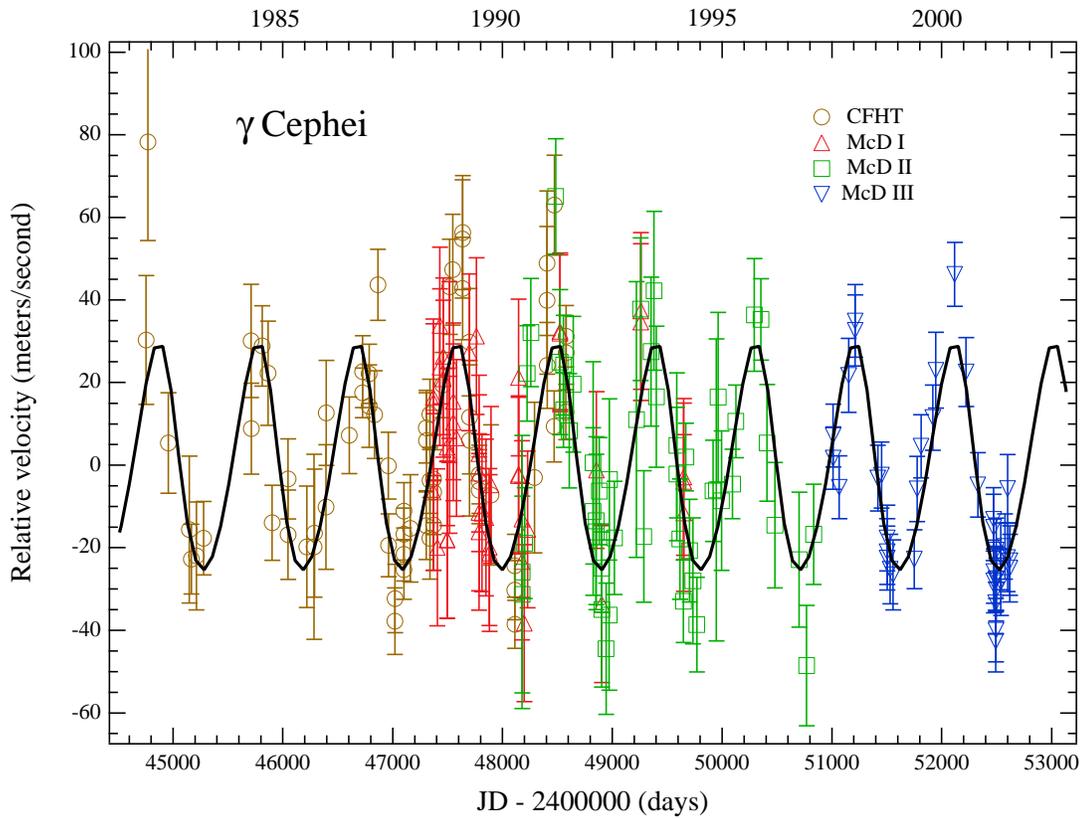,height=11cm,angle=0,clip=}
\end{center}
\caption{{Residual velocity measurements from  four data sets covering 21 years from CFHT and McDonald for $\gamma$ Cep after subtracting the stellar binary motion.  For the planet the orbital solution corresponds to a  minimum mass of 1.7 Jupiter in a 2.5 yr  (2.1 AU) eccentric orbit ($e=0.12$). \citep[Taken from][]{Hat03}}  
	\label{gCep92}
}
\end{figure}

\begin{figure}
\begin{center}
\epsfig{figure=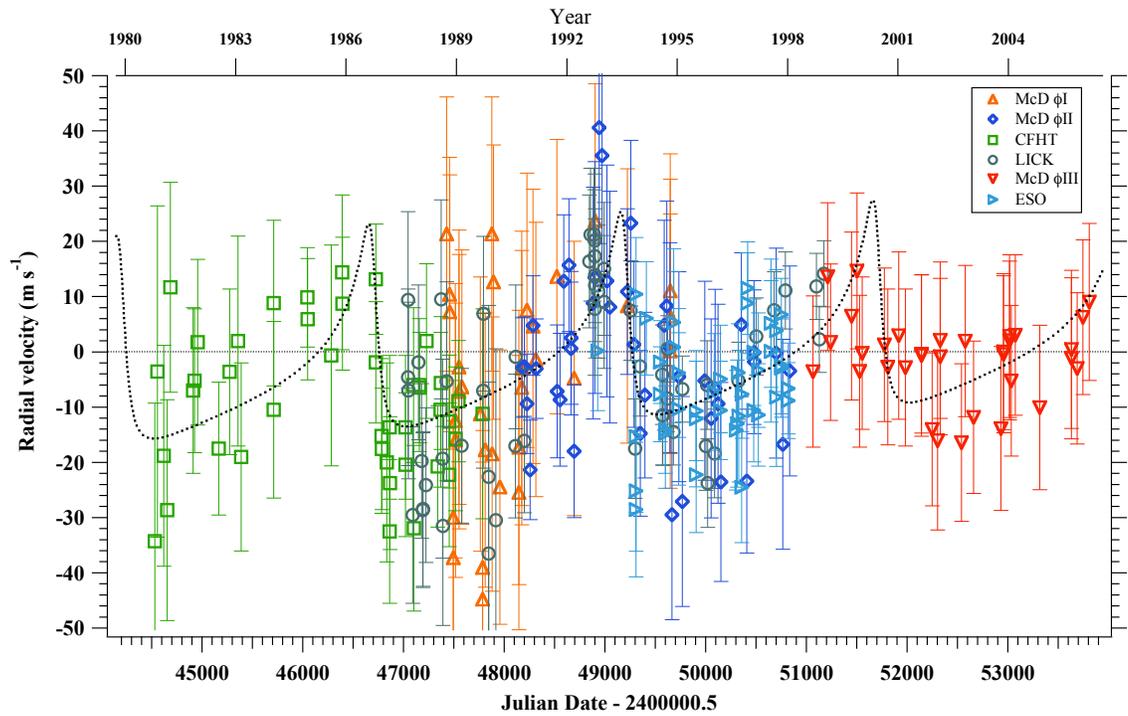,height=11.0cm,angle=0,clip=}
\end{center}
\caption{{The 26 yr radial velocity curve for $\epsilon$ Eri based on observations from CFHT, McDonald, Lick and ESO Observatories   \citep{Ben06}. The star is chromospherically and magnetically active and consequently a `noisy' target. A 6.9 yr periodicity seems to have persisted  with  an amplitude of  19 m s$^{-1}$ which corresponds to a minimum planetary mass of  0.9 Jupiter in an eccentric orbit ($e$=0.6) with a semi-major axis of 3.4 AU.   }  
	\label{eEri}
}
\end{figure}

\end{document}